# ARCH and GARCH Models
## vs.
## Martingale Volatility of Finance Market Returns

Joseph L. McCauley
Physics Department
University of Houston
Houston, Tx. 77204-5005
jmccauley@uh.edu

## Abstract

ARCH and GARCH models assume either i.i.d. or 'white noise' as is usual in regression analysis while assuming memory in a conditional mean square fluctuation with stationary increments. We will show that ARCH/GARCH is inconsistent with uncorrelated increments, violating the i.i.d. and 'white' assumptions and finance data and the efficient market hypothesis as well.

# 1. Introduction

In economics a point x at time t in a stochastic process x(t) is called a 'level' and an increment x(t,-T)=x(t)-x(t-T) is called a 'first difference'. Variables x that are logarithmic in a price variable are usual both in finance [1,2] and macroeconomics [3].

Given a stochastic process x(t), where in finance x(t)=ln(p(t)/$p_c$) [1,2] with $p_c$ is a reference price at time t [4], a stationary increment/difference [5,6,7,8] is one where x(t,-T)= ln(p(t)/p(t-T))=x(0,-T) in distribution, the difference is independent of the starting time t and depends only on the time lag T. To make this precise, with z=x(t,T)=y-x then the increment density is given by

$$f(z,t,t+T) = \iint dx dy f_2(y,t+T;x,t)\delta(z-y+x) \quad (1)$$

where $f_2$ is the 2-point density of returns, and increment stationarity means exactly that f(z,t,t+T)=f(z,0,T)= $\int dx f_2(x+z,T;x,0)$, nothing more and nothing less [2]. The 1-point density of returns or 'levels', in contrast, is given by $f_1(x,t) = \int dy f_2(y,t+T;x,t)$. If the condition (1) fails to hold, then the increments are nonstationary. Ito processes where (1) holds are nontrivial to construct, aside from the Wiener process [9].

In discussions of ARCH, GARCH in particular and regression analysis in general, an inadequate distinction is made between noise levels and noise increments. We've noted that the noise in regression eqns. must be interpreted as noise increments whether one assumes i.i.d. or 'white noise'. Therefore, what Engle [10] calls a 'variance', should be called a mean square fluctuation[1]. This will be made precise in the next section, where we will point out that "white noise" in econometrics means stationary noise increments with vanishing increment autocorrelations [9].

There are various volatility measures in practical use in finance theory [2]. The volatility measure chosen by Engle [10] is (in our language) the conditional mean square fluctuation $V(t,T) = \langle x^2(t,-T) \rangle_{cond}$. In a diffusive model (an Ito process) this would be given by

$$\langle x^2(t,-T) \rangle_{cond} = \int dy (y-x)^2 p_2(y,t+T|x,t) \quad (2)$$

where $p_2$ is the conditional density for the returns process $x(t)$.

In all that follows, we assume detrended data [1] and/or detrended stochastic models of levels x(t). With the choice x(0)=0 the process variance is given by $\sigma^2(t) = \langle x^2(t) \rangle$ where the process x(t) is then drift free noise. Only uncorrelated noise increments are of interest here. That is, we assume that the time lag T is sufficient that $\langle x(t,T)x(t,-T) \rangle \approx 0$. Next, we consider the basic regression models of volatility.

## 2. ARCH and GARCH Models

In regression analysis one begins with an eqn.

$$y(t) = \langle y(t) \rangle_{cond} + x(t,-T) \quad (3)$$

---

[1] We labeled a mean square fluctuation a 'variance' in [1], against this author's objection.

where t is the time, T is a time lag, and typically it's assumed that the conditional expectation is linear and time-lagged in y,

$$\langle y(t) \rangle_{cond} = \lambda_1 y(t-T) + \lambda_2 y(t-2T) + \ldots . \qquad (4)$$

The economists usually take T=1 period (as in one quarter of a year) but we avoid that restriction because it masks the fact that stationary increments cannot be treated as a stationary process if T is allowed to vary. The noise increment x(t,-T) is assumed to be "white", meaning in econometrics that (i) the increments are stationary, $x(t,-T) = x(0,-T)$ 'in distribution', (ii) $\langle x(0,-T) \rangle = 0$, (iii) the increments are uncorrelated $\langle x(0,-T)x(0,T) \rangle = 0$ (there is no reason to assume i.i.d. noise, lack of increment autocorrelations does the job [9]). Increment stationarity means that the mean square fluctuation is constant if T is held constant: $\langle x^2(t,-T)) \rangle = \langle x^2(0,-T)) \rangle$=constant for T=1, e.g. The reason that we denote the noise by "x" rather than "ε" will become clear when we introduce martingales in part 3.

ARCH models were proposed in 1982 [10] because of certain historical facts. The Black-Scholes model [11] was in its heyday, but the Black-Scholes model is nonvolatile: the detrended Gaussian returns model is the simplest martingale, and eqn. (2) for that model yields V(t,-T)=$\langle x^2(t,-T) \rangle = \sigma_1^2 T$ where $\sigma_1^2$ is constant. There is no volatility here because the dependence on the last observed point x at time t-T has disappeared. Volatility (2) requires an x-dependence, otherwise the conditional average V cannot fluctuate at all as t is increased.

The standard statement of an ARCH(1) process [12,13] is that with $\varepsilon_t / \langle y_t^2 \rangle$ assumed to be white noise, then

$$\langle y_t^2 \rangle_{cond} = \alpha + \omega y_{t-1}^2 \qquad (5a)$$

where the detrended returns are described by $\varepsilon_t = \ln(p(t)/p(t-1))$. Clearly, as has been pointed out recently [9], both the noise and the variable y here are not levels, they are both increments. Having made this clear, we now return to our standard notation for increments. We've pointed out elsewhere [1,2] that it's quite common, if mistaken, to regard the log increment $x(t)=\ln(p(t)/p(t-T))$ as a process, or level. In Ito calculus the levels obey stochastic differential eqns. and Fokker-Planck eqns., but the differences (except in the trivial case of a Wiener process) do not [2].

Historically, ARCH models were introduced to remedy the lack of volatility of the Gaussian returns model. The ARCH models were constructed with memory intentionally built into the mean square fluctuation [12]. Whether or not it was realized that the efficient market hypothesis (EMH) is violated is not clear because previous discussions of martingales as the EMH focused on simple averages and ignored pair correlations [14,15] (this is not entirely true, but Fama [15] stated the serial correlations of a martingale incorrectly [8]). We will show in part 3 below that the contradiction between ARCH and the EMH was probably masked by failing to distinguish between levels and differences in the noise.

The ARCH(1) model [10,12,13] is defined by the regression eqn.

$$\langle x^2(t,-T) \rangle_{cond} = \alpha + \omega x^2(t-T,-T) \qquad (5b)$$

with the assumption that the increments are stationary, are independent of t. In addition, the assumption was made that

$$x(t,-T) = z(T)\langle x^2(t,-T)\rangle_{cond}^{1/2} \qquad (6)$$

where z(T) was originally taken to be i.i.d. with zero mean and unit variance. It's adequate to assume that z(T) is uncorrelated [9] with zero mean and unit variance. The idea is that x(t,-T)=x(0,-T) 'in distribution' is the stationary noise in regression eqns. (3) if T is held fixed. So far, this is completely in the spirit of regression analysis: the noise is not assumed to have been discovered empirically, it's postulated in as simple a way as possible.

The unconditioned averages in ARCH(1) then obey

$$\langle x^2(t,-T)\rangle = \alpha + \omega\langle x^2(t-T,-T)\rangle. \qquad (7)$$

In regression analysis the assumption typically is that the increments are stationary. Stationary increments may have been inferred (rather, hypothesized) by 'eyeballing plots' of levels and differences [10,16,17], but were never verified, so far as we understand it [16], by an analysis based on constructing ensemble averages (ensemble averages are constructed from a single, long time series in ref. [1,2]). Accepting the assumption of stationary increments for now, we obtain

$$\langle x^2(t,-T)\rangle = \langle x^2(t-T,-T)\rangle = \langle x^2(0,-T)\rangle \qquad (8)$$

independent of t. This would yield

$$\langle x^2(0,-T)\rangle = \frac{\alpha(T)}{1-\omega(T)}. \qquad (9)$$

This is a relationship that can be checked, but that fact is masked by setting T=1 in regression analysis. We now show completely generally, without appeal to any particular dynamics, that ARCH(1) is completely inconsistent with 'white noise' (uncorrelated noise differences)

Increment autocorrelations are given by

$$2\langle x(t,-T)x(t,T)\rangle = \langle(x(t+T)-x(t-T))^2\rangle - \langle x^2(t,-T)\rangle - \langle x^2(t,T)\rangle. \qquad (10)$$

With stationary increments we obtain

$$2\langle x(0,-T)x(0,T)\rangle = \langle(x(0,2T)^2\rangle - 2\langle x^2(0,T)\rangle. \qquad (11)$$

The increment autocorrelations vanish iff. the levels variance is linear in the time [8], which then yields also that $\langle x^2(0,T)\rangle = T\langle x^2(0,1)\rangle$. Inserting this into (11), if we set T=0 then we obtain α=0. If T≠0 then we obtain ω=0. *This shows that ARCH(1) is inconsistent with stationary, uncorrelated increments.* The same conclusion will hold if the increments are nonstationary and uncorrelated. The reason for the contradiction is clear: uncorrelated increments guarantee a martingale x(t), and the martingale condition rules out memory at the level of simple averages and pair correlations [8]. ARCH models have finite memory built in at that level. The correct way to understand the ARCH models is that the memory is requires nonvanishing increment correlations. This violates the EMH and finance data as well [1,2]. Higher order ARCH models admit exactly the same interpretation.

The GARCH(1,1) model [12] is defined by

$$\langle x^2(t,-T) \rangle_{cond} = \alpha + \omega x^2(t-T,-T) + \zeta \langle x^2(t-T,-T) \rangle_{cond}. \quad (12)$$

If we again assume stationary increments then we obtain an analogous constant mean square fluctuation for fixed T. In this case 'white noise' would imply that α=0 and that $\omega + \zeta = 0$. With enough parameters the models are not falsifiable. There is no evidence for memory in observed finance market returns for T≥10 min. [1,2]. ARCH and GARCH models are only applicable to processes with correlated increments, and not to 'white noise' processes. In financial applications this requires lag time of T<10 min. in trading. Correlated increments occur for fractional Brownian motion, but not for efficient finance markets [1,2,8].

## Acknowledgement

JMC is grateful to both Duncan Foley and Barkley Rosser for helpful general comments and criticism via email, and especially to Søren Johansen for clarifying some of the assumptions made about noise in regression analysis in a related context [9].

## References

1. K.E. Bassler, J. L. McCauley, & G.H. Gunaratne, *Nonstationary Increments, Scaling Distributions, and Variable Diffusion Processes in Financial Markets*, PNAS **104**, 172287, 2007.

2. K.E. Bassler, G.H. Gunaratne, and J.L. McCauley, *Empirically Based modelling in finance and Beyond: Spurious Stylized Facts*, Int. Rev. Fin. An., 2008.

3. T.J. Sargent and N. Wallace, *J. Monetary Economics* 2, 169, 1976.

4. J. L. McCauley, K.E. Bassler, & G.H. Gunaratne, *On the Analysis of Time Series with Nonstationary Increments* in *Handbook of Complexity Research*, ed. B. Rosser, 2008.

5. R.L. Stratonovich. *Topics in the Theory of Random Noise*, Gordon & Breach: N.Y., tr. R. A. Silverman, 1963.

6. Mandelbrot & J. W. van Ness, *SIAM Rev.* 10, 2, 422,1968.

7. P. Embrechts and M. Maejima, *Selfsimilar Processes*, Princeton University Press, Princeton, 2002.

8. J.L. McCauley, K.E. Bassler, & G.H. Gunaratne, *Martingales, Detrending Data, and the Efficient Market Hypothesis, Physica* **A387**, 202, 2008.

9. J.L. McCauley, K.E. Bassler, and G.H. Gunaratne, *'Integration I(d)' of Nonstationary Time Series: When are increments of noise processes stationary,* preprint, 2008.

10. R.F. Engle, *Econometrica* **50**, nr. 4, 987, 1982.

11. F. Black and M. Scholes, *J. Political Economy* **8**, 637, 1973.

12. Arch and Garch Models, http://www.quantlet.com/mdstat/scripts/sfe/html/sfenode66.html

13. The Royal Swedish Academy of Sciences, *Time Series Econometrics: Cointergration and Autoregressive Conditional Heteroskedasticity,*

http://www.kva.se/KVA_Root/files/newspics/DOC_2003108143127_50163615451_ecoadv03.pdf, 8 Oct. 2003.

14. B. Mandelbrot, *J. Business* **39**, 242, 1966.

15. E. Fama, *J. Finance* **25**, 383-417, 1970.

16. Søren Johansen, email discussion, 2008.

17. K. Juselius & R. MacDonald, *International Parity Relations Between Germany and the United States: A Joint Modelling Approach*, preprint, 2003.